\documentclass[preprint,3p,twocolumn,sort&compress]{elsarticle}

\usepackage{atlasphysics}
\usepackage{amsmath}
\usepackage{subfigure}
\usepackage{units}
\usepackage{booktabs}
\usepackage{lineno}
\usepackage{placeins}
\usepackage[hyperindex,breaklinks]{hyperref} 

\newcommand{\proton}{p}
\newcommand{\bquark}{b}
\newcommand{\cquark}{c}
\newcommand{\topquark}{t}
\newcommand{\antitop}{\bar{\topquark}}
\renewcommand{\Wboson}{W}
\newcommand{\Wt}{\Wboson\!\topquark}
\newcommand{\pp}{\proton\proton}
\newcommand{\Wjets}{\Wboson\!\text{+jets}}
\newcommand{\WjetsHF}{\Wboson\!\text{+heavy-flavour}}
\newcommand{\WjetsLF}{\Wboson\!\text{+light-jets}}
\newcommand{\Zjets}{\Zboson\text{+jets}}
\renewcommand{\ttbar}{\topquark\antitop}

\newcommand{\Pt}{p_{\text{T}}}
\newcommand{\kt}{k_{t}}
\newcommand{\Etmiss}{E_{\text{T}}^{\text{miss}}}
\newcommand{\mtW}{m_{\text{T}}^{\Wboson}}

\usepackage{preprintcover}

\PreprintCoverPaperTitle{Search for s-channel single top-quark production in proton--proton collisions 
at $\mathsf{\sqrt{s}\!=\!8}$ TeV with the ATLAS detector}
\PreprintIdNumber{CERN-PH-EP-2014-224}

\PreprintCoverAbstract{This Letter presents a search at the LHC for $s$-channel single top-quark production 
in proton--proton collisions at a centre-of-mass energy of $\unit[8]{TeV}$. The analyzed data set 
was recorded by the ATLAS detector and corresponds to an integrated luminosity of 
$\unit[20.3]{fb^{-1}}$. Selected events contain one charged lepton, large missing transverse 
momentum and exactly two $b$-tagged jets. A multivariate event classifier based on boosted 
decision trees is developed to discriminate $s$-channel single top-quark events from the main 
background contributions. The signal extraction is based on a binned maximum-likelihood fit 
of the output classifier distribution. The analysis leads to an upper limit on the $s$-channel 
single top-quark production cross-section of $\unit[14.6]{pb}$ at the 95\% confidence level. The 
fit gives a cross-section of $\sigma_s=\unit[5.0\pm4.3]{pb}$, consistent with the Standard 
Model expectation.}

\PreprintJournalName{Physics Letters B}

\begin{document}
\begin{frontmatter}

\title{Search for $s$-channel single top-quark production in proton--proton collisions at $\sqrt{s}=8$~TeV 
with the ATLAS detector}
\author{The ATLAS Collaboration}

\begin{abstract}
This Letter presents a search at the LHC for $s$-channel single top-quark production in 
proton--proton collisions at a centre-of-mass energy of $\unit[8]{TeV}$. The analzed data set 
was recorded by the ATLAS detector and corresponds to an integrated luminosity of 
$\unit[20.3]{fb^{-1}}$. Selected events contain one charged lepton, large missing transverse 
momentum and exactly two $b$-tagged jets. A multivariate event classifier based on boosted 
decision trees is developed to discriminate $s$-channel single top-quark events from the main 
background contributions. The signal extraction is based on a binned maximum-likelihood fit 
of the output classifier distribution. The analysis leads to an upper limit on the $s$-channel 
single top-quark production cross-section of $\unit[14.6]{pb}$ at the 95\% confidence level. The 
fit gives a cross-section of $\sigma_s=\unit[5.0\pm4.3]{pb}$, consistent with the Standard 
Model expectation.
\end{abstract}


\end{frontmatter}

\section{Introduction}
\label{sec:introduction}

At hadron colliders, top-quarks are produced predominantly in pairs ($\ttbar$) via the 
strong interaction but can also be produced singly through the electroweak interaction. 
At leading order in perturbation theory, there are three possible single top-quark 
production modes: an exchange of a virtual $\Wboson$ boson either in the $t$-channel 
or in the $s$-channel, or the associated production of a top-quark and a $\Wboson$ boson. 
In the $s$-channel, the exchange of a time-like $\Wboson$ boson produces a top-quark 
and a bottom-quark in the final state~\cite{Cortese:1991}. Independent measurements of 
these three modes are of great interest since different impacts on their production rates 
are predicted by the various proposed scenarios beyond the Standard Model~\cite{Tait:2000}. 

Single top-quark production was first observed at the Tevatron in proton--antiproton 
collisions by the CDF and D0 Collaborations in measurements of the combined $s$- and 
$t$-channel production~\cite{CDF:2009,Dzero:2009}. The observation of the $s$-channel 
process alone through the combination of the CDF~\cite{CDF:2014} and D0~\cite{Dzero:2013} 
cross-section measurements was reported recently~\cite{CDF_Dzero:2014}. At the Large Hadron 
Collider (LHC), the production cross-section of single tops in proton--proton ($\pp$) 
collisions was measured both in the $t$-channel and in association with a $\Wboson$ boson by 
the CMS~\cite{CMS_tchan,CMS_tchan_2,CMS_Wt,CMS_Wt_2} and ATLAS~\cite{ATLAS_tchan,ATLAS_tchan_2,ATLAS_Wt} 
Collaborations. Measuring the $s$-channel process is more difficult due to a much smaller 
signal-to-background ratio, this production mode being disadvantaged at the LHC due 
to the need for a sea antiquark in the initial state. 

Calculations at approximate next-to-next-to-leading-order (NNLO) precision in QCD are available
for the production cross-section of single tops~\cite{Kidonakis_tchan,Kidonakis_Wt,Kidonakis_schan}.
These approximate NNLO results include the contributions due to the next-to-next-to-leading-logarithm 
(NNLL) resummation of soft-gluon bremsstrahlung. For the $s$-channel process, the total inclusive 
cross-section for $\pp$ collisions at a centre-of-mass energy of $\unit[8]{TeV}$ is predicted to be 
$\sigma_s\!=\!\unit[5.61\pm0.22]{pb}$. This assumes a top-quark mass of $\unit[172.5]{\GeV}$ and 
uses the MSTW2008~\cite{PDF_Martin} NNLO set of parton distribution functions (PDFs). The quoted 
uncertainty includes the QCD scale uncertainty and the correlated PDF--$\alpha_{\mathrm{s}}$ 
uncertainty. The cross-sections calculated at approximate NNLO for the dominant $t$-channel and $Wt$ 
processes are $\sigma_t=\unit[87.8^{+3.4}_{-1.9}]{pb}$~\cite{Kidonakis_tchan} and 
$\sigma_{Wt}=\unit[22.4\pm1.5]{pb}$~\cite{Kidonakis_Wt}, respectively.

This Letter presents a search for $s$-channel single top-quark production in $\pp$ collisions 
at $\sqrt{s}=8$~TeV carried out at the LHC by the ATLAS Collaboration. Only leptonic decay modes 
of the top-quark giving an electron or a muon are considered and the signal is extracted from a 
likelihood fit to the distribution of a multivariate discriminant. For all reported results, 
the electron and muon channels are merged independently of the lepton charge in order to measure 
the combined production cross-section of top-quarks and top-antiquarks.

\section{Data sample}
\label{sec:data}

The analysis is performed on $\pp$ collision data collected by the ATLAS detector~\cite{ATLAS_detector} 
in 2012 at a centre-of-mass energy of $\unit[8]{TeV}$ and corresponding to an integrated luminosity 
of $\unit[20.3\pm0.6]{fb^{-1}}$~\cite{ATLAS_lumi}.

The ATLAS detector is a multipurpose apparatus consisting of a precise tracking system, calorimeters 
and a muon spectrometer. The inner tracking system contains a high-granularity silicon pixel detector, 
a silicon microstrip tracker and a straw-tube transition radiation tracker. This system exploits a 
$\unit[2]{T}$ axial magnetic field and provides charged-particle tracking in the 
pseudorapidity\footnote{ATLAS uses a right-handed coordinate system with its origin at the nominal 
interaction point in the centre of the detector and the $z$-axis along the beam pipe. The pseudorapidity 
is defined in terms of the polar angle $\theta$ as $\eta\!=\!-\ln\tan(\theta/2)$. Distances in 
($\eta$,$\phi$) space are defined by $\Delta R=\sqrt{(\Delta\eta)^2+(\Delta\phi)^2},$ where $\phi$ 
denotes the azimuthal angle around the beam pipe.} region $|\eta|<2.5$. The central calorimeter system 
(barrel) is divided into a liquid-argon electromagnetic sampling calorimeter with high granularity 
and a hadron calorimeter consisting of iron and scintillator tiles. The former covers a range 
of $|\eta|<1.47$, while the latter extends to $|\eta|=1.7$. The two endcap regions are equipped with 
liquid-argon calorimeters for electromagnetic and hadronic energy measurements up to $|\eta|=4.9$.  
The muon spectrometer consists of three large superconducting toroids with eight coils each, and 
includes precision tracking chambers for momentum measurements up to $|\eta|=2.7$ and fast trigger 
chambers covering the range $|\eta|<2.4$. A three-level trigger system, consisting of custom-made 
hardware followed by two software-based selections, is used to yield a recorded event rate of about 
$\unit[400]{Hz}$.

Events recorded by single-electron or single-muon triggers under stable beam conditions and with good 
quality data from all detector subsystems are selected for the analysis. The chosen trigger 
thresholds give a constant efficiency for lepton candidates passing the offline selections. 
Each triggered event includes on average about $\unit[20]{}$ additional $\pp$ collisions (pile-up) 
from the same bunch crossing.

\section{Event reconstruction and selection}
\label{sec:selection}

The experimental signature of the searched-for $s$-channel single top events consists of a 
single isolated electron or muon, significant missing transverse momentum due to the undetected 
neutrino from the $\Wboson$ boson leptonic decay, and exactly two $b$-tagged jets, one of 
them being associated with the top-quark decay. Events in which the $\Wboson$ boson decays to a 
$\tau$ lepton are also included if the $\tau$ lepton decays subsequently to an electron or a muon. 

Electron candidates are reconstructed from an isolated electromagnetic energy deposit matched to 
an inner detector track and passing tight identification requirements~\cite{ATLAS_electrons}. The 
candidates are required to have a transverse energy greater than $\unit[30]{\GeV}$ and a pseudorapidity 
$\abseta<2.47$, excluding the barrel--endcap transition regions of the electromagnetic calorimeters, 
$1.37<\abseta<1.52$. Muon candidates are reconstructed by combining tracks reconstructed in both the 
inner detector and the muon spectrometer~\cite{ATLAS_muons, ATLAS_muons_2}. The candidates are required 
to have a transverse momentum $\Pt$ greater than $\unit[30]{\GeV}$ and $\abseta<2.5$. The electron 
and muon candidates must also fulfill isolation requirements, as described in Ref.~\cite{Top_isolation}, 
in order to reduce contributions from misidentified jets, non-prompt leptons from heavy-flavour decays 
and non-prompt electrons from photon conversions.

Jets are reconstructed, using the anti-$\kt$ algorithm~\cite{Antikt} with a radius parameter of
$0.4$, from calorimeter energy clusters calibrated with the local cluster weighting method~\cite{ATLAS_jets_1}. 
Jets are calibrated using an energy- and $\eta$-dependent simulation-based calibration scheme with 
in situ corrections based on data~\cite{ATLAS_jets_2}. In this analysis, jets with $\Pt>\unit[30]{\GeV}$ 
and $\abseta<2.5$ are selected. Jets likely to originate from the hadronization of $\bquark$-quarks are 
identified using a multivariate discriminant which makes use of track impact parameters and reconstructed 
secondary vertices~\cite{ATLAS_btag_2}. Jets are defined to be $\bquark$-tagged (hereinafter referred to 
as $\bquark$-jets) if the discriminant value is above a threshold corresponding to a $\bquark$-tagging 
efficiency of 70\% for simulated $\ttbar$ events~\cite{ATLAS_btag_3}; the associated rejection factors 
against light-quark and charm-quark jets are about $140$ and $5$, respectively~\cite{ATLAS_btag_2}.

The missing transverse momentum, with magnitude $\Etmiss$, is reconstructed from the vector sum of 
all clusters of energy deposits in the calorimeters calibrated at the electromagnetic scale; the 
contributions associated with the reconstructed electrons and jets are corrected 
to the corresponding energy scales~\cite{ATLAS_met_1}. Contributions from muons are also taken into account using their 
measured transverse momentum. The $\Etmiss$ is a measurement of the escaping neutrinos but also 
includes energy losses due to detector inefficiencies. In this analysis, $\Etmiss$ is required to 
be greater than $\unit[35]{\GeV}$. 

Events are required to have at least one reconstructed primary vertex with at least five associated 
tracks with $\Pt>\unit[400]{\MeV}$, and no jets failing reconstruction quality requirements. To enhance 
the signal content of the sample of events containing a single isolated lepton and exactly two 
$\bquark$-jets, the events are in addition not allowed to contain any other jets with a transverse 
momentum greater than $\unit[25]{\GeV}$. In addition, the $\Wboson$ boson transverse 
mass,\footnote{The $\Wboson$ boson transverse mass is computed from the lepton transverse momentum
$\Pt^{\ell}$, the missing transverse momentum $\Etmiss$ and their difference in azimuthal angle as
$\mtW=\sqrt{2\Etmiss\Pt^{\ell}\left(1-\cos(\Delta\phi(\Etmiss,\,\Pt^{\ell}))\right)}\,$.}
$\mtW$, must be greater than $\unit[50]{\GeV}$ in order to reduce the multijet background contribution.

The kinematics of the neutrino and of the top-quark are reconstructed. The transverse momentum of the
neutrino is assumed to be given by the $x$- and $y$-components of the missing transverse momentum, 
while the unmeasured longitudinal component of the neutrino momentum is extracted by imposing a 
$\Wboson$ boson mass constraint on the lepton--neutrino system. The latter leads to a quadratic 
equation in the longitudinal momentum component of the neutrino, $p_{\mathit{z}}^{\nu}$. If there 
are two real solutions the ambiguity is resolved by choosing the one with the smallest 
$|p_{\mathit{z}}^{\nu}|$. For complex solutions due to the non-perfect resolution of the missing 
transverse momentum measurement, the imaginary component is eliminated by modifying 
$\Etmiss$ in such a way that the transverse mass of the $\Wboson$ candidate fulfills the $\Wboson$ 
mass constraint, while preserving the direction of the missing transverse momentum. The kinematics 
of two top-quark candidates are then reconstructed by combining the four-momenta of the lepton and 
neutrino with each of the two $\bquark$-jets (called leading and sub-leading jets according to their 
ranking in $\Pt$). The top-quark candidate with an invariant mass closest to $\unit[172.5]{\GeV}$ 
defines the best candidate. To improve the signal-to-background ratio of the selected events, events 
for which the transverse momentum of the $\bquark$-jet corresponding to the best top-quark candidate 
is lower than $\unit[50]{\GeV}$ are rejected. 

After all selection requirements, 16031 events are selected from the analyzed data sample.

\section{Simulated samples}
\label{sec:simulations}

Simulated event samples are used in the analysis for comparison with the data as well as to 
evaluate signal and background contributions and uncertainties. The simulation samples for 
the single top and $\ttbar$ processes are produced with the next-to-leading-order (NLO)
matrix-element generator {\sc Powheg-Box} (v1\_r2129)~\cite{Powheg} using the CT10 PDF 
set~\cite{PDF_Lai}. The generator is interfaced to {\sc Pythia}~(6.42)~\cite{Pythia} for 
parton showering, hadronization and underlying-event modelling with parameters set to the 
values of the Perugia 2011C tune~\cite{PerugiaTune}.

For the evaluation of the uncertainty due to the generator and parton shower modelling, 
additional samples for the $s$-channel, $\Wt$ and $\ttbar$ processes are produced using 
the {\sc MC@NLO} generator~(4.03)~\cite{MCatNLO_1,MCatNLO_2,MCatNLO_3,MCatNLO_4} 
interfaced to {\sc Herwig}~(6.52)~\cite{Herwig_1,Herwig_2} 
for parton showering and {\sc Jimmy}~(4.31)~\cite{Jimmy} for the underlying-event modelling 
with the ATLAS AUET2 tune~\cite{AUETTune} and the CT10 PDFs. For single-top $t$-channel
production, the {\sc MadGraph5\_aMC@NLO}~(2.0)~\cite{aMCatNLO} generator also interfaced with 
{\sc Herwig} and {\sc Jimmy} is employed. To estimate the uncertainty coming from the amount 
of initial-state and final-state radiation, samples of $\ttbar$ events are produced with the 
leading-order (LO) {\sc AcerMC}~(3.8)~\cite{AcerMC} generator using the CTEQ6L1 PDF 
set~\cite{PDF_Pumplin}. The generator is interfaced to {\sc Pythia}~(6.42) and the parameters 
controlling the radiation emission are varied. The parameter settings are constrained by jet
activity measurements in $\ttbar$ production~\cite{ATLAS_ISRFSR_2}. The impact of scale variations 
on the signal events is studied using $s$-channel samples generated with {\sc Powheg-Box}, 
interfaced to {\sc Pythia}, with values of the factorization and renormalization scales being 
increased or decreased by a factor of two; this scale variation also causes the jet multiplicity 
to vary. 

All samples involving top-quark production are generated using a top-quark mass of $\unit[172.5]{\GeV}$. 
The single top samples are normalized to the approximate NNLO cross-section 
predictions~\cite{Kidonakis_tchan,Kidonakis_Wt,Kidonakis_schan} presented in Section~\ref{sec:introduction}.
The $\ttbar$ samples are normalized to the cross-section, $\sigma_{\ttbar}=\unit[253^{+13}_{-15}]{pb}$, 
calculated at NNLO in QCD including resummation of NNLL soft-gluon terms with 
Top{\scriptsize ++}2.0~\cite{Cacciari:2011,Baernreuther:2012,Czakon:2012a,Czakon:2012b,Czakon:2013,Czakon:2011}.

Event samples simulating the production of $\Wboson$ and $\Zboson$ bosons in association with 
jets ($\Wjets$ and $\Zjets$) are produced using the LO multiparton generator {\sc Alpgen}~(2.14)~\cite{Alpgen}, 
with the CTEQ6L1 set of PDFs, and interfaced to {\sc Pythia}~(6.42). The {\sc Alpgen} matrix elements 
include diagrams with up to five partons. $\Wjets$ samples with only light-quark jets as well as samples 
with heavy-flavour quarks ($Wb\bar{b}$+jets, $Wc\bar{c}$+jets, $Wc$+jets) are produced separately. To 
evaluate the modelling uncertainty related to $\Wjets$ production, alternative samples are produced using 
the LO multiparton generator {\sc Sherpa}~(1.4.1)~\cite{Sherpa} with the CT10 PDF set. The diboson 
processes ($WW$, $WZ$, $ZZ$) are simulated using the {\sc Herwig}~(6.52) and {\sc Jimmy} generators
with the ATLAS AUET2 tune and the CTEQ6L1 PDF set. The single-boson and diboson samples are normalized 
to their inclusive production cross-sections calculated at NNLO~\cite{Anastasiou:2003} or 
NLO~\cite{Campbell:2011} precision.

All generated events are passed through the full ATLAS detector simulation~\cite{ATLAS_FullSim} 
based on {\sc Geant}4~\cite{GEANT4}, or through a faster simulation making use of parameterized
calorimeter showers~\cite{ATLAS_FastSim}. Minimum-bias events, generated with {\sc Pythia}~(8.1)~\cite{Pythia8}, 
are overlaid to simulate the pile-up effects from additional $\pp$ collisions in the same and nearby 
bunch crossings. All simulated events are then processed using the same reconstruction and analysis 
chain as for data.

\section{Background estimation}
\label{sec:backgrounds}

The predicted event yields for the signal and backgrounds after the selection described in 
Section~\ref{sec:selection} are summarized in Table~\ref{tab:yields} with a comparison of the 
total expectation to the data.

The main background contribution is due to $\ttbar$ events, which account for 67\% of the total 
expected yield. The two next-largest backgrounds come from $t$-channel single top and 
$\WjetsHF$ production, each process accounting for 12\% of the total expectation. The rather 
large predicted cross-sections for these processes which involve one lepton and one or two 
$\bquark$-jets, coupled to a non-negligible misidentification of jets originating from lighter 
quark flavours, explain their significant contributions. The individual contributions associated 
with the $Wt$, $\WjetsLF$, $\Zjets$ and diboson background processes are all lower than 3\%. 

The estimated contribution due to multijet events is about 2\%.  Multijet events pass the signal 
selection if in addition to two reconstructed and $\bquark$-tagged jets an extra jet is misidentified 
as an isolated lepton, or a non-prompt lepton appears to be isolated (both referred to as a fake lepton). 
This background is estimated from data using the matrix method~\cite{MatrixMethod}. In this approach, 
a set of equations is solved, which relates the observed sample composition in terms of selected leptons 
of two different categories, loose and tight, to its true composition in terms of prompt (real) and fake 
leptons. The tight category corresponds to the signal selection mentioned in Section~\ref{sec:selection}, 
while the isolation requirements are removed to define the loose category, which also has a loosened 
identification quality in the electron case. The real and fake efficiencies are measured using dedicated 
samples of data enriched in real and fake isolated leptons. An uncertainty of 50\% on the multijet event 
yield is evaluated from comparisons with alternative procedures for the efficiency extraction and 
alternative criteria for the selection of the real- and fake-enriched samples. 

\begin{table}[t]
\centering
\begin{tabular}{lr@{$\,\pm\,$}l}
\hline
\hline
Process                       & \multicolumn{2}{c}{Event yield} \\
\hline
Single top $s$-channel        & 457     & 50              \\
Single top $t$-channel, $\Wt$ & 2270    & 240             \\
$\ttbar$                      & 10200   & 1600            \\
$\Wjets$                      & 1900    & 1200            \\
$\Zjets$, diboson             & 108     & 68              \\
Multijet                      & 230     & 120             \\[1ex]
Total expectation             & 15200   & 2000            \\
Data                          & \multicolumn{2}{c}{16031} \\
\hline
\hline
\end{tabular}  
\caption{Predicted and observed event yields after all selection requirements. The quoted 
errors include all systematic uncertainties described in Section~\ref{sec:systematics} 
added in quadrature with the simulation statistical uncertainties. For the multijet background, 
the data-driven estimate is reported.}
\label{tab:yields}
\end{table}

The modelling of the backgrounds is validated by comparing the expected event distributions to the 
data in two control regions. The first control sample is defined by requiring both jets to pass a 
relaxed $\bquark$-tagging selection with an efficiency of 80\%, while requiring at least one 
of them to fail the tighter 70\% $\bquark$-tagging selection used for the signal selection. The 
dominant contribution, which comes from the $\Wjets$ process, is expected to represent 56\% of the 
events while the expected signal-to-background ratio is lower than 1\%. The second control sample is 
selected by requiring two jets that are not $\bquark$-tagged in addition to the two signal $\bquark$-jets. 
This control region is dominated by $\ttbar$ events (around 90\% of the total) with a negligible 
$s$-channel single top contribution. For both control selections, the threshold on the $\Wboson$ boson 
transverse mass is lowered to $\unit[30]{\GeV}$ and the selection on the $\Pt$ of the $\bquark$-jet 
associated with the best top-quark candidate is not applied. Good overall agreement between data and 
expectation is observed in both control regions.

\section{Multivariate analysis}
\label{sec:analysis}

After event selection, the estimated signal-to-background ratio is about 3\%. To improve the
separation of $s$-channel single top events from backgrounds, several kinematic and topological 
variables are combined into one discriminant through a multivariate classification technique. 
The event classification is based on boosted decision trees (BDTs)~\cite{BDT} in the TMVA 
framework~\cite{TMVA}.

\begin{figure*}[t]
\centering
\subfigure[\label{subfig:dPhiTop2Bjet}]{\includegraphics[width=0.46\textwidth]{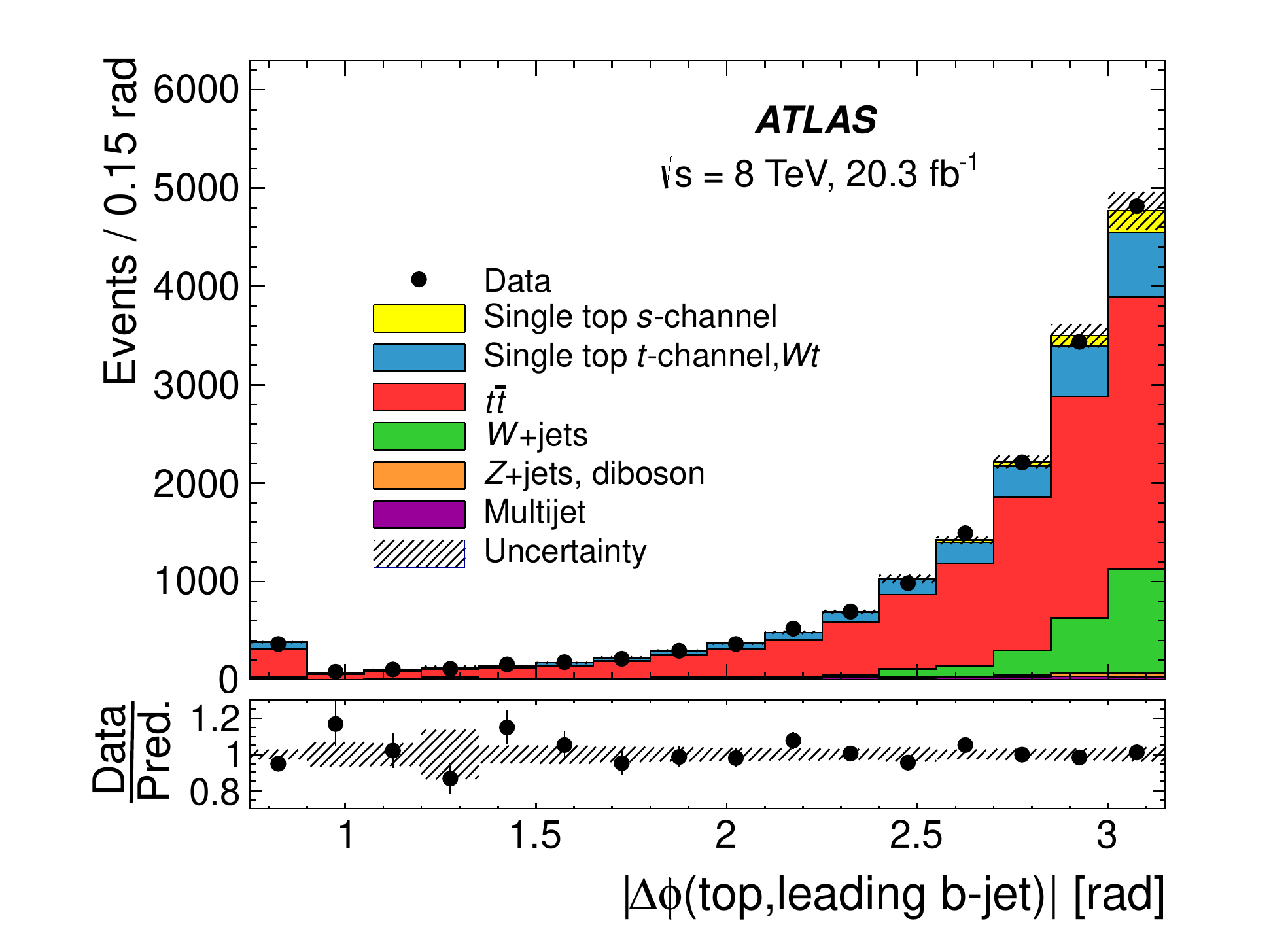}}
\subfigure[\label{subfig:HtLeptonMet}]{\includegraphics[width=0.46\textwidth]{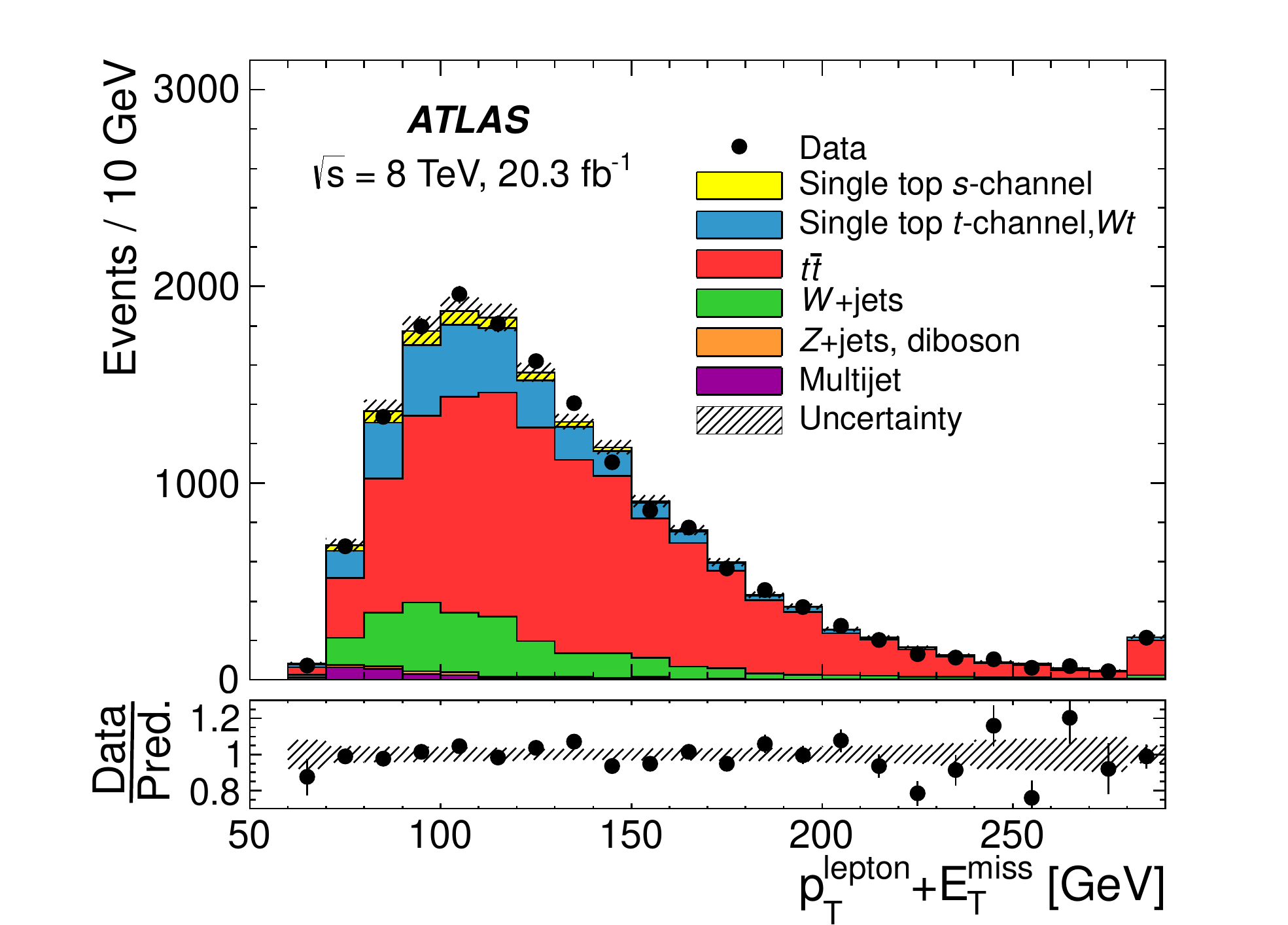}}
\subfigure[\label{subfig:WtransMass}]{\includegraphics[width=0.46\textwidth]{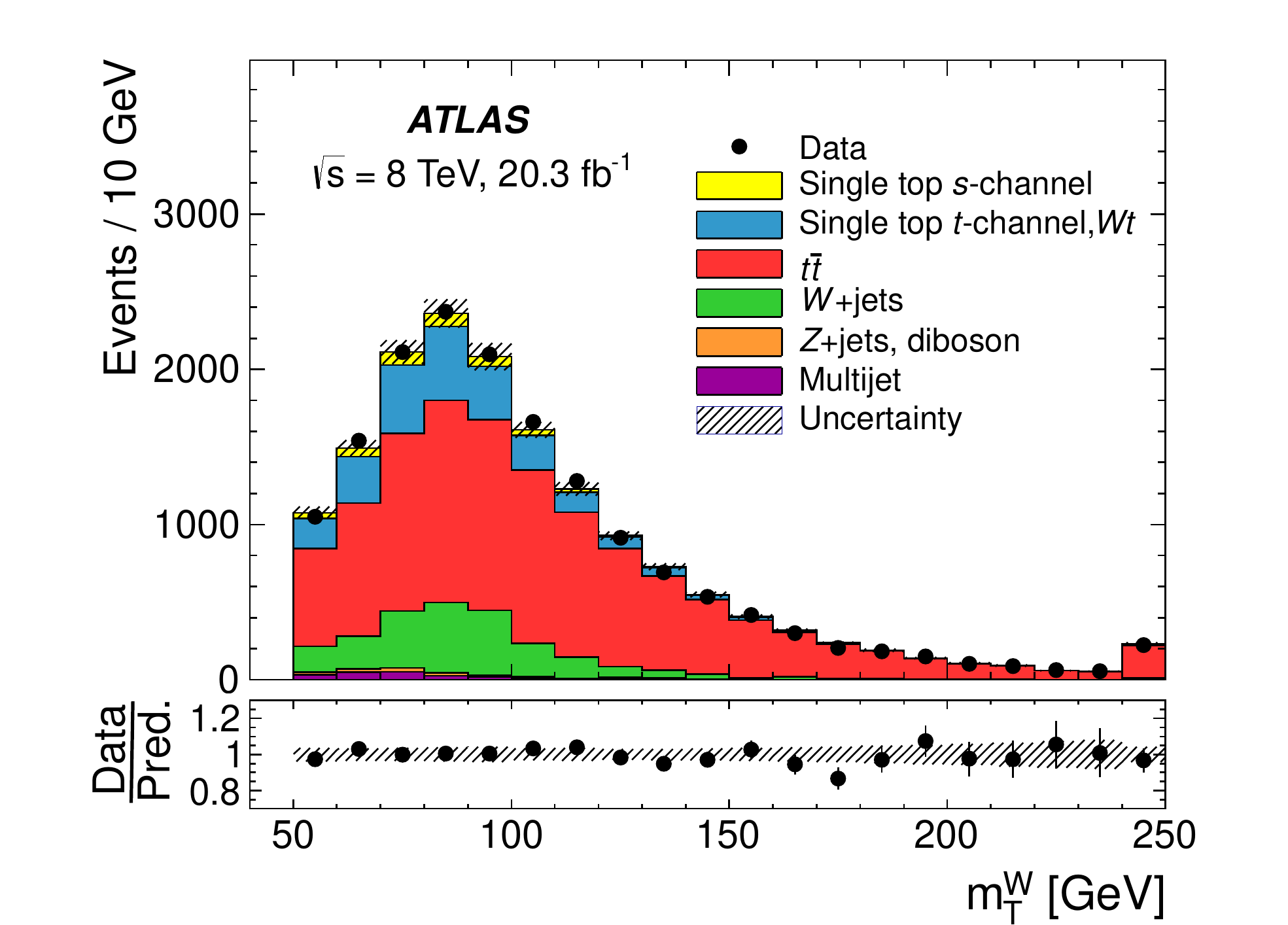}}
\subfigure[\label{subfig:dEtaLeptonBjet}]{\includegraphics[width=0.46\textwidth]{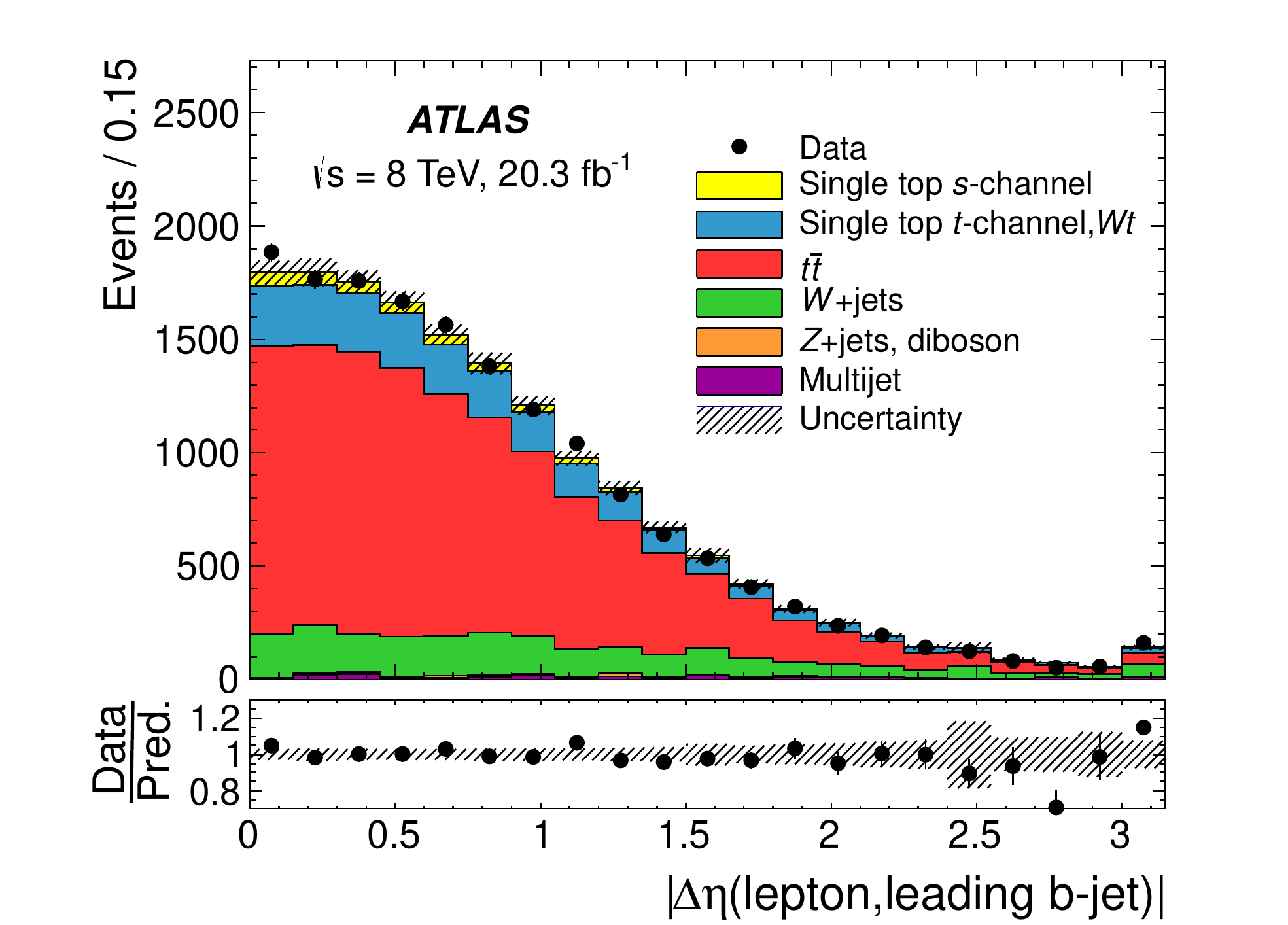}}
\caption
{Observed and predicted distributions in the signal region of four of the most discriminating 
variables used in the BDT classifier
\subref{subfig:dPhiTop2Bjet} azimuthal angle between the leading $\bquark$-jet and the top-quark 
candidate reconstructed with the sub-leading $\bquark$-jet, 
\subref{subfig:HtLeptonMet} scalar sum of the lepton momentum and the missing transverse momentum,
\subref{subfig:WtransMass} $\Wboson$ boson transverse mass, and
\subref{subfig:dEtaLeptonBjet} difference in pseudorapidity between the lepton and the leading 
$\bquark$-jet. The simulated distributions are normalized to the result of the binned maximum-likelihood 
fit described in Section~\ref{sec:xsection}. The hatched bands represent the normalization uncertainty
of all processes after the fit added in quadrature with the simulation statistical uncertainty. 
The lower panels show the ratio of data to prediction. The first (last) bin of the histograms includes 
underflows (overflows).}
\label{fig:BDT_variables}
\end{figure*}

Nineteen variables with significant separation power are used as input to the BDT classifier. The number 
of variables is chosen in order to maximize the expected signal significance and all of the retained 
variables are well modelled by simulation in the signal region and in the two control regions defined 
in Section~\ref{sec:backgrounds}. The two most discriminating variables are the differences in azimuthal 
angle ($|\Delta\phi|$) between the leading (sub-leading) $\bquark$-jet and the top-quark candidate reconstructed 
with the sub-leading (leading) $\bquark$-jet. The four next-most important variables are the scalar sum 
of the lepton $\Pt$ and $\Etmiss$, $\Etmiss$, the $\Wboson$ boson transverse mass, and the lepton $\Pt$. 
The other variables, which have a smaller discriminating power, are (grouped according to their type): 
({\it i}) the differences in pseudorapidity ($|\Delta\eta|$) between the lepton and the leading (sub-leading) 
$\bquark$-jet, the $|\Delta\eta|$ between the leading $\bquark$-jet and the top-quark candidate reconstructed with the 
sub-leading $\bquark$-jet, the $|\Delta\eta|$ between the reconstructed neutrino and the $\bquark$-jet not 
associated with the best top-quark candidate, ({\it ii}) the $|\Delta\phi|$ between the lepton and the missing 
transverse momentum, the $|\Delta\phi|$ between the two $\bquark$-jets, ({\it iii}) the scalar sum of the $\Pt$ 
of all objects (lepton, $\bquark$-jets, neutrino), the magnitude of the vector sum of the $\Pt$ of the two 
$\bquark$-jets, the invariant mass of the two $\bquark$-jets, and ({\it iv}) the cosine of the angle ($\cos\vartheta$) 
between the missing transverse momentum and the sub-leading $\bquark$-jet, the $\cos\vartheta$ between the 
lepton in the $\Wboson$ rest frame and the $\Wboson$ boson in the top-quark candidate rest frame reconstructed with 
the leading $\bquark$-jet, the $\cos\vartheta$ between the lepton in the top-quark candidate rest frame and the top-quark 
candidate reconstructed with the leading (sub-leading) $\bquark$-jet in the centre-of-mass frame defined with all objects.

The multivariate classifier is trained using simulated events selected with relaxed criteria: the threshold on the 
$\Wboson$ boson transverse mass is lowered to $\unit[30]{\GeV}$ and the selection on the transverse momentum 
of the $\bquark$-jet associated with the best top-quark candidate is not applied. The classifier response is 
tuned with the gradient boosting algorithm\ \cite{TMVA} and with only the $\ttbar$ and $\WjetsHF$ background 
contributions included in the training phase.

\begin{figure}[t]
\centering
\includegraphics[width=0.52\textwidth]{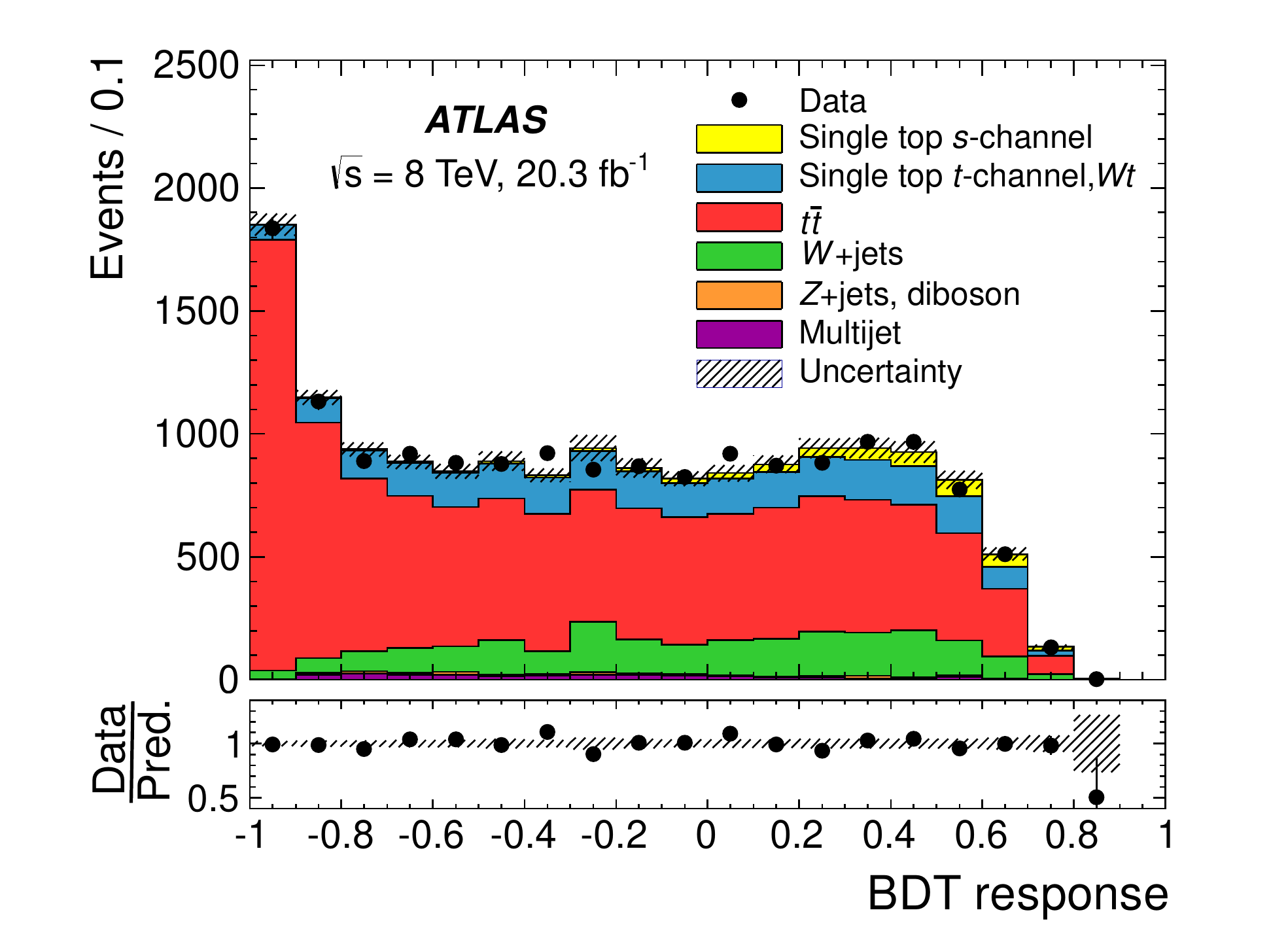}
\caption
{Observed and predicted distributions of the BDT response. The simulated distributions are normalized 
to the result of the binned maximum-likelihood fit. The hatched band represents the normalization 
uncertainty of all processes after the fit added in quadrature with the simulation statistical 
uncertainty. The lower panel shows the ratio of data to prediction.}
\label{fig:BDT_response}
\end{figure}

\begin{figure}[t]
\centering
\includegraphics[width=0.46\textwidth]{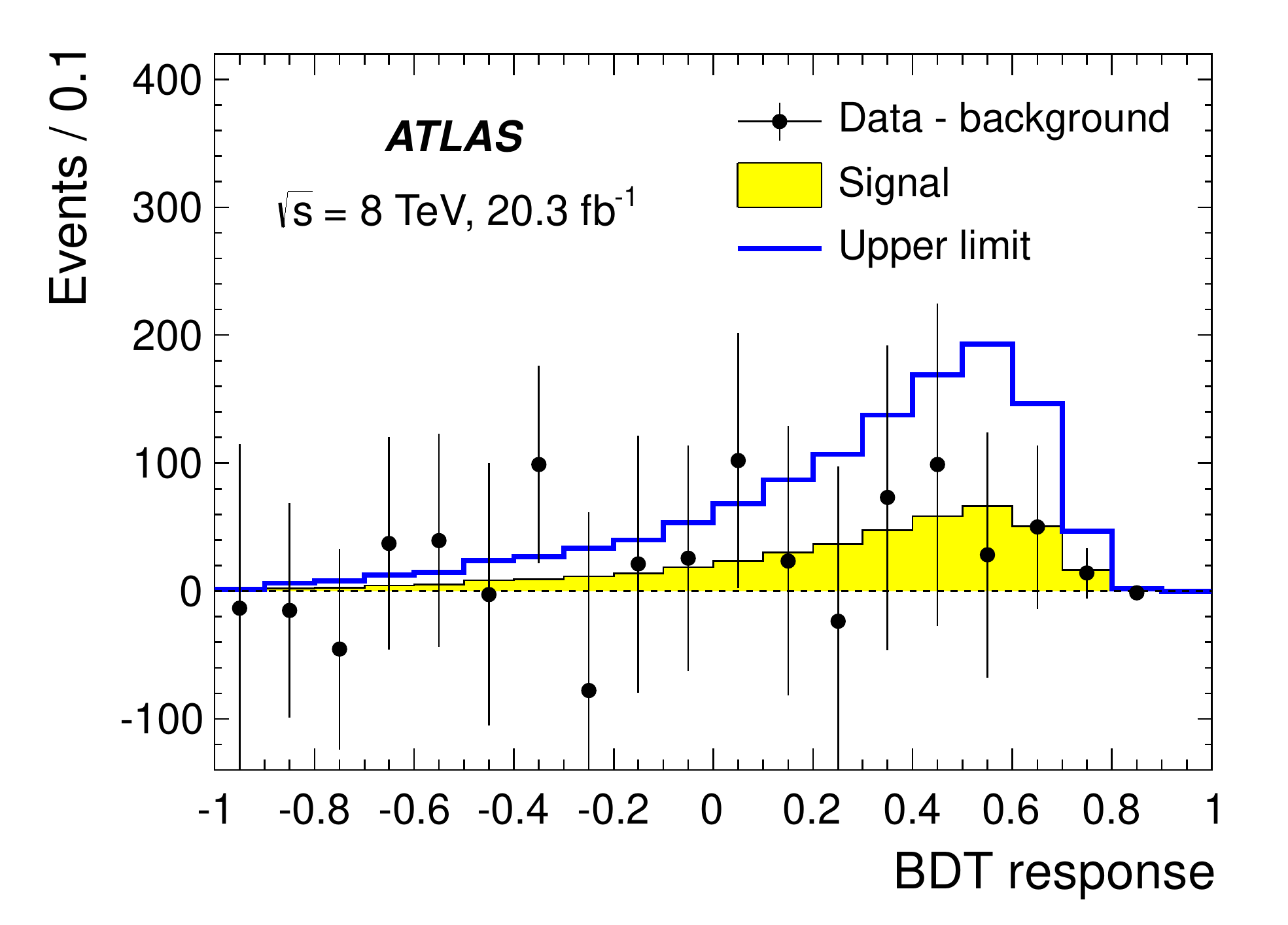}
\caption
{Observed and predicted (filled histogram) signal distributions of the BDT response after background 
subtraction. The signal and subtracted simulated background contributions are normalized to the result 
of the binned maximum-likelihood fit. The error bars represent the statistical uncertainty of the data 
added in quadrature with the simulation statistical uncertainties and with the systematic uncertainties 
on the background normalizations described in Section~\ref{sec:systematics}. The empty histogram 
represents the predicted signal distribution renormalized to the observed cross-section upper limit 
reported in Section~\ref{sec:results}.}
\label{fig:BDT_result}
\end{figure}

Fig.~\ref{fig:BDT_variables} shows the comparison between data and prediction for four of the main 
discriminating variables used as input to the BDT classifier. The observed and expected distributions of 
the BDT response are presented in Fig.~\ref{fig:BDT_response}. The observed and expected signal distributions 
of the BDT response are shown in Fig.~\ref{fig:BDT_result}; the former is obtained by subtracting the background 
contributions from the data. The simulated distributions are normalized to the result of the binned 
maximum-likelihood fit performed to extract the signal content from the observed distribution (see 
Section~\ref{sec:xsection}). The $t$-channel and $Wt$ single top contributions as well as the $\Zjets$ and 
diboson components are merged.

\section{Systematic uncertainties}
\label{sec:systematics}

Several sources of systematic uncertainty affect the signal acceptance and the background normalizations, 
as well as the shape of the BDT distribution. The impact on the distribution of using simulation samples 
of limited size is also taken into account. The various sources of systematic uncertainties considered 
when extracting the $s$-channel single top signal are described below.

Systematic uncertainties on the reconstruction and energy calibration of jets, electrons and muons 
are propagated in the analysis through variations in the modelling of the detector response. For the 
jets, the main source of uncertainty is the energy scale, evaluated using a combination of in situ 
techniques~\cite{ATLAS_jets_2}. Other jet-related uncertainty sources are the modelling of the energy 
resolution~\cite{ATLAS_jets_3} and reconstruction efficiency~\cite{ATLAS_jets_2}, and the modelling of 
the tagging efficiencies of $\bquark$-jets, $\cquark$-jets and light-quark jets~\cite{ATLAS_btag_2,ATLAS_btag_3}. 
Uncertainties related to the leptons come from trigger, identification and isolation efficiencies, as 
well as from the energy scale and resolution~\cite{ATLAS_electrons,ATLAS_muons_2}. The uncertainties from 
the energy scale and resolution corrections applied to leptons and jets are propagated to the computation 
of the missing transverse momentum. The scale and resolution uncertainties due to soft jets ($\Pt<\unit[20]{\GeV}$) and to 
contributions of calorimeter energy deposits not associated with any reconstructed objects are also 
considered and evaluated independently (uncertainties referred to as $\Etmiss$ scale and resolution).

The luminosity measurement is calibrated using dedicated beam-separation scans and the same methodology 
as that detailed in Ref.~\cite{ATLAS_lumi}. The resulting uncertainty on the integrated luminosity is 2.8\%.

Systematic uncertainties on the simulation of the signal and background processes ($\ttbar$, $t$-channel 
single top, $Wt$ and $\Wjets$) are also taken into account in the analysis. They include contributions 
from the modelling of the hard process, parton showers and hadronization and of initial-state and final-state 
radiation (ISR/FSR). The uncertainty due to the choice of renormalization and factorization scales is also 
considered for the signal. These various uncertainties are estimated by comparing simulation samples produced, 
as described in Section~\ref{sec:simulations}, with different generators, different shower models and different 
settings for the amount of additional radiation ($\ttbar$) or for the scales ($s$-channel).

The systematic uncertainties associated with the PDFs are taken into account for all simulated samples. 
They are assessed according to the PDF4LHC prescription~\cite{PDF4LHC} and using the MSTW2008~\cite{PDF_Martin}, 
CT10~\cite{PDF_Lai}, and NNPDF2.1~\cite{PDF_Ball} sets.

Other sources of uncertainty are related to the background normalizations, which are taken from the theory 
predictions with their associated uncertainties, except for the multijet production for which a data-driven 
normalization is used. Theoretical uncertainties of 6\% and 5\% are considered for $\ttbar$ and for the 
combination of the $t$-channel and $Wt$ single top contributions, respectively. For $\ttbar$, the PDF and 
$\alpha_\mathrm{s}$ uncertainties, calculated using the PDF4LHC prescription~\cite{PDF4LHC} with the MSTW2008 
NNLO~\cite{PDF_Martin,PDF_Martin_2}, CT10 NNLO~\cite{PDF_Lai,PDF_Gao} and NNPDF2.3 5f FFN~\cite{PDF_Ball} 
PDF sets, are added in quadrature to the QCD scale uncertainty. The normalization uncertainties for the 
$\Wjets$ and for the combined $\Zjets$ and diboson backgrounds are estimated to be 60\%; they include an 
estimate of the uncertainty on the heavy-flavour fraction of 50\%, added in quadrature with a theory 
uncertainty of 5\% for the inclusive process, and a Berends scaling uncertainty of 24\% for each extra 
jet~ \cite{Ellis:1985,Berends:1990}. For multijet events, a normalization uncertainty of 50\% is estimated 
(see Section~\ref{sec:backgrounds}). The theoretical uncertainty of 4\% on the $s$-channel single-top 
cross-section is considered when evaluating the signal significance.

\section{Signal extraction}
\label{sec:xsection}

The signal contribution to the selected sample of data is extracted by performing a binned
maximum-likelihood fit to the BDT output distribution. The likelihood function is given by 
the product of Poisson probability terms for the individual distribution bins (see 
Ref.~\cite{ATLAS_tchan}), combined with the product of Gaussian functions to constrain the 
background rates to their predicted values within the associated uncertainties. The rates of the 
$\ttbar$, $\Wjets$, and combined $t$-channel\,/\,$Wt$ and $\Zjets$\,/\,diboson backgrounds 
are thus fitted within their theoretical uncertainties described in Section~\ref{sec:systematics}; 
since the multijet component is obtained from the data it is not allowed to vary and is fixed to 
its data-driven estimate. The production cross-section is derived from the adjusted signal rate 
for which the logarithm of the likelihood function reaches its maximum.

The systematic uncertainties on the measurement are determined using pseudo-experiments involving 
variations of the signal and background rates and of the shape of the BDT distributions, due to all the 
sources of uncertainty described in Section~\ref{sec:systematics}. The uncertainties due to 
the limited size of the data and simulation samples are also assessed via pseudo-experiments implementing
statistical fluctuations. The impact of an individual source of uncertainty is evaluated by running 
dedicated pseudo-experiments with only the corresponding variation included. The total uncertainty 
is evaluated from pseudo-experiments including all variations simultaneously.

The sensitivity to the $s$-channel single top signal is determined by testing both the background-only 
and the signal-plus-background hypotheses via the generation of dedicated sets of pseudo-experiments; 
in the second case, the signal yield is set to the approximate NNLO prediction. A test statistic, defined 
as the logarithm of the ratio combining the maximum-likelihood estimates of the two alternative hypotheses, 
is computed for each pseudo-experiment. The test statistic distribution is used to derive the $p$-value of 
the background-only ensemble test, which is then interpreted in terms of signal significance. A cross-section 
upper limit is extracted using the CLs procedure~\cite{Junk:1999,Read:2002}, which is based on the $p$-values 
calculated for both the background-only and the signal-plus-background ensemble tests. 

\section{Results}
\label{sec:results}

Using the frequentist approach presented above, the observed (expected) significance of the $s$-channel 
single top measurement is found to be 1.3 (1.4) standard deviations and an observed (expected) upper 
limit on the production cross-section of $\unit[14.6]{pb}$ ($\unit[15.7]{pb}$, $\unit[9.4]{pb}$) is set 
at the 95\% confidence level. The two quoted expected upper limits correspond to the signal-plus-background 
and background-only hypotheses, respectively. The simulated BDT distribution of the signal events 
renormalized to the observed upper limit is presented in Fig.~\ref{fig:BDT_result} for comparison with 
the signal distribution extracted from the data.

\begin{table}[t!]
\centering
\begin{tabular}{ll}
\hline
\hline
Source                         & $\Delta\sigma/\sigma$ [\%] \\
\hline
Data statistics                & $\pm$35  \\
Simulation statistics          & $\pm$29  \\[1ex]
$\Etmiss$ scale                & $\pm$54  \\
$\Etmiss$ resolution           & $+$0/$-$3 \\
Jet energy scale               & $\pm$39  \\
Jet energy resolution          & $\pm$5   \\
Jet tagging efficiencies       & $\pm$4   \\
Jet reconstruction efficiency  & $<$ 1    \\
Lepton energy scale/resolution & $<$ 1    \\
Lepton efficiencies            & $+$2/$-$1 \\[1ex]
Signal modelling and scale    & $\pm$11  \\
$\ttbar$ modelling             & $\pm$6   \\
$\Wjets$ shape modelling       & $\pm$8   \\
ISR/FSR                        & $\pm$3   \\
PDF                            & $<$ 1    \\[1ex]
Background normalization       & $\pm$7   \\
Multijet normalization         & $\pm$12  \\[1ex]
Integrated luminosity          & $\pm$5   \\[1ex]
Total systematic               & $\pm$80  \\
Total                          & $\pm$87  \\
\hline
\hline
\end{tabular}
\caption{Contributions of the sources of statistical and systematic uncertainty to the total
uncertainty on the measured cross-section. They are given in percent.}
\label{tab:breakdown}
\end{table}

The fitted value of the cross-section is found to be: 
$\sigma_s=\unit[5.0\pm1.7\text{\,(stat.)}\pm4.0\text{\,(syst.)}]{pb}=\unit[5.0\pm4.3]{pb}$; 
this measurement should be compared with the cross-section of $\unit[5.61\pm0.22]{pb}$ calculated 
at approximate NNLO. Table~\ref{tab:breakdown} summarizes the various contributions to the 
measurement uncertainty. The largest systematic uncertainties arise from the scale of the missing 
transverse momentum (54\%) and of the jet energy (39\%). These large uncertainties are due to a strongly 
distorted shape of the BDT distribution obtained when adding bin-per-bin the systematic variations 
affecting the signal and background contributions. The limited size of the data sample and the simulated 
samples also contribute significantly to the final uncertainty (35\% and 29\%, respectively). Smaller 
uncertainties are due to the background normalizations (12\% for the data-driven multijet contribution 
and 7\% from theory for the simulated backgrounds) and process modelling (11\%, 8\% and 6\% for signal, 
$\Wjets$ and $\ttbar$, respectively). Other modelling and instrumental effects play only a minor role.

\section{Summary}
\label{sec:summary}

This Letter presents a search for $s$-channel single top production at the LHC from the $pp$ 
collision data sample of $\unit[20.3]{fb^{-1}}$ recorded by the ATLAS detector at a centre-of-mass 
energy of $\unit[8]{TeV}$. A multivariate analysis, based on boosted decision trees, is carried out to 
discriminate signal from background events. The observed signal significance is 1.3 standard deviations, 
the expected sensitivity being 1.4 standard deviations. The observed upper limit on the $s$-channel single 
top cross-section is $\unit[14.6]{pb}$ at the 95\% confidence level. The evaluated production cross-section 
is $\sigma_s=\unit[5.0\pm1.7\text{\,(stat.)}\pm4.0\text{\,(syst.)}]{pb}=\unit[5.0\pm4.3]{pb}$ and it is
consistent with the Standard Model prediction.

\section*{Acknowledgements}


We thank CERN for the very successful operation of the LHC, as well as
the support staff from our institutions without whom ATLAS could not
be operated efficiently.

We acknowledge the support of ANPCyT, Argentina; YerPhI, Armenia; ARC,
Australia; BMWFW and FWF, Austria; ANAS, Azerbaijan; SSTC, Belarus;
CNPq and FAPESP, Brazil; NSERC, NRC and CFI, Canada; CERN; CONICYT,
Chile; CAS, MOST and NSFC, China; COLCIENCIAS, Colombia; MSMT CR, MPO
CR and VSC CR, Czech Republic; DNRF, DNSRC and Lundbeck Foundation,
Denmark; EPLANET, ERC and NSRF, European Union; IN2P3-CNRS,
CEA-DSM/IRFU, France; GNSF, Georgia; BMBF, DFG, HGF, MPG and AvH
Foundation, Germany; GSRT and NSRF, Greece; ISF, MINERVA, GIF, I-CORE
and Benoziyo Center, Israel; INFN, Italy; MEXT and JSPS, Japan; CNRST,
Morocco; FOM and NWO, Netherlands; BRF and RCN, Norway; MNiSW and NCN,
Poland; GRICES and FCT, Portugal; MNE/IFA, Romania; MES of Russia and
ROSATOM, Russian Federation; JINR; MSTD, Serbia; MSSR, Slovakia; ARRS
and MIZ\v{S}, Slovenia; DST/NRF, South Africa; MINECO, Spain; SRC and
Wallenberg Foundation, Sweden; SER, SNSF and Cantons of Bern and
Geneva, Switzerland; NSC, Taiwan; TAEK, Turkey; STFC, the Royal
Society and Leverhulme Trust, United Kingdom; DOE and NSF, United
States of America.

The crucial computing support from all WLCG partners is acknowledged
gratefully, in particular from CERN and the ATLAS Tier-1 facilities at
TRIUMF (Canada), NDGF (Denmark, Norway, Sweden), CC-IN2P3 (France),
KIT/GridKA (Germany), INFN-CNAF (Italy), NL-T1 (Netherlands), PIC
(Spain), ASGC (Taiwan), RAL (UK) and BNL (USA) and in the Tier-2
facilities worldwide.

\section*{References}

\bibliographystyle{atlasBibStyleWoTitle}
\bibliography{references}

\onecolumn
\clearpage
\begin{flushleft}
{\Large The ATLAS Collaboration}

\bigskip

G.~Aad$^{\rm 85}$,
B.~Abbott$^{\rm 113}$,
J.~Abdallah$^{\rm 152}$,
S.~Abdel~Khalek$^{\rm 117}$,
O.~Abdinov$^{\rm 11}$,
R.~Aben$^{\rm 107}$,
B.~Abi$^{\rm 114}$,
M.~Abolins$^{\rm 90}$,
O.S.~AbouZeid$^{\rm 159}$,
H.~Abramowicz$^{\rm 154}$
% ...
\bigskip
\\
% ...
$^{11}$ Institute of Physics, Azerbaijan Academy of Sciences, Baku, Azerbaijan\\
% ...
$^{85}$ CPPM, Aix-Marseille Universit{\'e} and CNRS/IN2P3, Marseille, France\\
% ...
$^{90}$ Department of Physics and Astronomy, Michigan State University, East Lansing MI, United States of America\\
% ...
$^{107}$ Nikhef National Institute for Subatomic Physics and University of Amsterdam, Amsterdam, Netherlands\\
% ...
$^{113}$ Homer L. Dodge Department of Physics and Astronomy, University of Oklahoma, Norman OK, United States of America\\
$^{114}$ Department of Physics, Oklahoma State University, Stillwater OK, United States of America\\
% ...
$^{117}$ LAL, Universit{\'e} Paris-Sud and CNRS/IN2P3, Orsay, France\\
% ...
$^{152}$ Institute of Physics, Academia Sinica, Taipei, Taiwan\\
% ...
$^{154}$ Raymond and Beverly Sackler School of Physics and Astronomy, Tel Aviv University, Tel Aviv, Israel\\
% ...
$^{159}$ Department of Physics, University of Toronto, Toronto ON, Canada\\
% ...
\end{flushleft}


\end{document}